\documentclass[aps,superscriptaddress,a4paper,twocolumn,showpacs,floatfix]{revtex4}
\usepackage[dvips]{graphicx}
\usepackage{amssymb}

\def\Tss{\ensuremath{T_\mathrm{ss}}}
\def\Tspin{\ensuremath{T_\mathrm{s}}}
\def\Tmelt{\ensuremath{T_\mathrm{m}}}
\def\Tstar{\ensuremath{T^*}}
\def\gammaSV{\ensuremath{\gamma_\mathrm{SV}}}
\def\gammaLV{\ensuremath{\gamma_\mathrm{LV}}}
\def\gammaSL{\ensuremath{\gamma_\mathrm{SL}}}

\begin{document}
\title{Physics of Solid and Liquid Alkali Halide Surfaces Near
          the Melting Point}

\author{T. Zykova-Timan}
\affiliation{International School for Advanced Studies (SISSA), and INFM
                DEMOCRITOS National Simulation Center,\\
                via Beirut 2-4, I-34014 Trieste, Italy}

\author{D. Ceresoli}
\affiliation{International School for Advanced Studies (SISSA), and INFM
                DEMOCRITOS National Simulation Center,\\
                via Beirut 2-4, I-34014 Trieste, Italy}

\author{U. Tartaglino}
\affiliation{International School for Advanced Studies (SISSA), and INFM
                DEMOCRITOS National Simulation Center,\\
                via Beirut 2-4, I-34014 Trieste, Italy}
\affiliation{IFF, FZ-J\"ulich, 52425 J\"ulich, Germany}

\author{E. Tosatti}
\affiliation{International School for Advanced Studies (SISSA), and INFM
                DEMOCRITOS National Simulation Center,\\
                via Beirut 2-4, I-34014 Trieste, Italy}
\affiliation{International Centre for Theoretical Physics
                (ICTP), P.O.Box 586, I-34014 Trieste, Italy}

\begin{abstract}
This paper presents a broad theoretical and simulation study of the high
temperature behavior of crystalline alkali halide surfaces typified by
NaCl(100), of the liquid NaCl surface near freezing, and of the very unusual
partial wetting of the solid surface by the melt. Simulations are conducted
using two-body rigid ion BMHFT potentials, with full
treatment of long-range Coulomb forces. After a preliminary check
of the description of bulk NaCl provided by these potentials, which seems
generally good even at the melting point, we carry out a new investigation of
solid and liquid surfaces. Solid NaCl(100) is found in this model
to be very anharmonic and yet
exceptionally stable when hot. It is predicted by a thermodynamic integration
calculation of the surface free energy that NaCl(100) should be a well
ordered, non-melting surface, metastable even well above the melting point.
By contrast, the simulated liquid NaCl surface is found to exhibit large
thermal fluctuations and no layering order. In spite of that, it is shown to
possess a relatively large surface free energy. The latter is traced to a
surface entropy deficit, reflecting some kind of surface short range order.
Finally, the solid-liquid interface free energy is
derived through Young's equation from direct simulation of partial wetting
of NaCl(100) by a liquid droplet.
It is concluded that three elements, namely
the exceptional anharmonic stability of the solid (100) surface, the
molecular short range order at the liquid surface, and the
costly solid liquid interface, all conspire to cause the
anomalously poor wetting of the (100) surface by its own melt in
the BMHFT model of NaCl --  and most likely also in real alkali halide
surfaces.
\end{abstract}
\maketitle
\newpage
\clearpage
\section{Introduction}\label{sec:introduction}
Attention is increasing toward adhesion and wetting, the structure and physics of
solid-liquid interfaces especially at high temperatures, and the structure
of liquid surfaces, particularly of complex and molecular systems. In order
to gain more insight into these problems, there is a strong need for good
case studies, to use as well-understood starting points.

One easy starting point is to study the contact of a liquid
with \emph{its own} solid, a clear situation where there will be no
ambiguity of physical description, no uncertainty in chemical composition,
no segregation phenomena, all of them complications present in the study of
contact between different substances.
Contact of a liquid with the surface of its own solid usually materializes
spontaneously at high temperature. Most solid surfaces are known to wet
themselves spontaneously with an atomically thin film of melt, when their
temperature $T$ is brought close enough to the melting point \Tmelt\ of the
solid. The phenomenon whereby the thickness $l(T)$ of the liquid film
diverges continuously (and critically) as $T \to \Tmelt$, is commonly
referred to as (complete) surface melting~\cite{vander,physrep}.

Surface melting is indeed a very natural thing to happen, because it
corresponds to complete wetting of a solid substrate by the same identical
substance, only in liquid form.
Another name for surface melting, better suited for
the fluid community~\cite{dietrich,dietrich1}, could be complete interfacial wetting.
Due to surface melting, a solid with free surfaces cannot generally be
overheated above its thermodynamical bulk \Tmelt. The free energy barrier
for the passage from solid to liquid generally requires nucleation and
implies hysteresis. Although this is not generally observable for reasons exposed below,
it should be theoretically possible to overheat a solid, at
least insofar as one can exclude the presence of defects that could act as
nucleation centers.

In the melting phase transition, the solid surface itself represents a (nearly)
ubiquitous defect that continuously nucleates the liquid, thus usually
preventing overheating of the solid. A generic crystal surface simply does not
remain solid at high temperature, and spontaneously wets itself with a liquid
film of increasing thickness very close to the melting point. The
thermodynamical condition for surface melting, or complete interfacial
wetting, to occur, is that the solid-vapor (SV) interface should turn itself
spontaneously into the solid-liquid (SL) plus liquid-vapor (LV) interface pair,
or:
\begin{equation}\label{eq1}
    \gammaSV = \gammaSL + \gammaLV ,
\end{equation}
where the $\gamma$'s are interface free energies at the triple point.

There are a number of known exceptions to this behavior. On some solid
surfaces the liquid film does begin to form upon heating, but its thickness
levels off to a finite value instead of diverging as $T \to\, \Tmelt$
(so-called blocked or incomplete~\cite{incomplete} surface melting). More
remarkably, some other solid surfaces remain dry and fully crystalline up to the 
bulk triple
point. This \emph{surface non-melting} phenomenon, originally discovered in 
molecular
dynamics simulations of Au(111)~\cite{carnevali} and independently observed
experimentally in Pb(111)~\cite{Pb111}, takes place at the close-packed faces
of several metals, such as Al(111)~\cite{Al111}.

Thermodynamically, surface non-melting will occur if there is a
free energy loss in converting the SV interface into the SL plus LV
interfaces pair, namely
\begin{equation}\label{eq:young1}
    \gammaSV < \gammaSL + \gammaLV.
\end{equation}

\begin{figure*}[!ht]
     \includegraphics[width=0.45\textwidth]{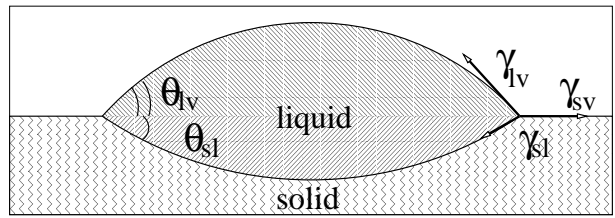}
     \caption{ \label{scheme} Sketch of the liquid drop partially wetting its own solid, showing the balance of the forces acting at the interfaces.}
\end{figure*}

In that case, the liquid will wet the solid at best {\em incompletely}.
A liquid droplet deposited in full equilibrium on the solid surface will
not spontaneously spread. It will settle instead in a metastable partial
wetting geometry such as that of Fig.~\ref{scheme}. The metastable droplet
is characterized by two angles $\theta_{SL},\theta_{LV}$, and by two curvature 
radii (not shown)
$R_{SL}, R_{LV}$ of the solid-liquid and liquid-vapor interfaces,
approximately obeying the generalized Young equations~\cite{nozieres}:
\begin{eqnarray}\label{mech}
     \gammaSV &=& \gammaSL\cos \theta_{SL} + \gammaLV \cos\theta_{LV} \\
     R_{LV}\sin\theta_{LV} &=& R_{SL}\sin\theta_{SL}
\end{eqnarray}
The temporary settling of a metastable liquid droplet on the surface of
the same solid substance, schematically depicted in Fig.~\ref{scheme} and
discussed by Nozi\'eres~\cite{nozieres}, was  demonstrated in simulation
in Ref.~\cite{ditolla95} for Al/Al(111), but not verified experimentally
yet~\cite{note1}.

Here we wish to move on from elemental to molecular systems. Alkali halides
are our natural prototype choice, because they represent a well defined
class of substances whose liquids do not wet their own solid, and
because they have otherwise long been studied experimentally and theoretically.
Molten salts and their surfaces were extensively investigated
by macroscopic techniques~\cite{croxton}. A partial wetting angle $\theta_{LV}$= 
48$^\circ$
is known for liquid NaCl on NaCl(100) at the melting point \Tmelt\ = 1074 K
(and similar results also hold for other alkali halides)~\cite{mutaftschiev75,mutaftschiev97}.
Such a large angle underlines a strikingly poor wetting of the liquid onto its
own solid, large by comparison with other known cases. For example,
$\theta_{LV} \sim$15$^\circ$ is observed for liquid Pb/Pb(111)~\cite{frenken},
or $ \sim 19^\circ $ is obtained by simulation of Au/Au(111)~\cite{ditolla95}.
Should liquid surface layering be, as was the case in metals, the culprit
in NaCl too, the layering magnitude and its effect should indeed be
exceptionally strong, and amenable to experimental verification.
Several simulations on alkali halide liquids suggest instead that there is no layering
whatsoever~\cite{heyes,madden}.
A second possible mechanism leading to surface non-melting may
arise from van der Waals forces. Whenever the melt is optically more dense
than the solid, the so-called Hamaker constant $H$ governing the effective
SL-LV interface interaction $H/l^2$ may turn negative. The resulting
attraction impedes complete melting, as is the case for valence
semiconductors such as silicon~\cite{physrep}.
However, liquid NaCl is 26\% less
dense than the solid, and here the Hamaker constant is certainly
positive~\cite{articleforTosi}.

That leaves the question of explaining the exceptionally poor wetting of
NaCl(100) by liquid NaCl completely open. In order to shed light on this
question and on the underlying physics, we undertook extensive simulation
studies of the molten NaCl surface, and also of the NaCl(100) solid surface,
at and close to the melting point.

The plan of the rest of this paper is as follows. We will introduce first the 
simulation
methods and the potentials used. Subsequently we will proceed to a careful
characterization of all bulk properties of this model. The bulk solid and
the bulk liquid will be simulated, and the results compared to experimental
data and to previous theory and simulations. The bulk zero-pressure melting 
temperature,
very close to the triple point temperature, will be extracted for this
model potential by direct simulation of the solid-liquid coexistence. Next,
extensive slab simulations will be used to obtain a quantitative
description of the solid (100) surface  near and above bulk melting,
and of the liquid surface in the same temperature range.
The solid surface free energy
$\gamma_{SV}$ will be calculated by thermodynamic integration. That
of the liquid surface $\gamma_{LV}$  will be calculated by means of the
Kirkwood-Buff virial formula. The two will be compared
and discussed. An effective harmonic calculation will be implemented
showing  the importance of anharmonicity in stabilizing the high
temperature solid surface. A  modified calculation of the liquid surface
free energy  will be introduced to  understand the poor temperature
dependence of $\gamma_{LV}$, eventually explained in terms of surface short range
order. Finally the solid-liquid interface free energy $\gamma_{SL}$ will
be calculated via Young's equation from a direct simulation of partial
wetting of a liquid droplet on the solid surface at the melting point.
The resulting large value of $\gammaSL$ is connected to the large density
jump. In the concluding discussion, it is argued that all three separate
physical mechanisms that conspire to give rise to poor wetting, by lowering
$\gamma_{SV}$ and simultaneously  raising $\gamma_{LV}$ and $\gamma_{SL}$
eventually stem from charge order and charge neutrality of this ionic system.

\section{Hamiltonian, and Simulation Method}\label{sec:hamiltonian}
NaCl was modeled by the classic pairwise Born-Mayer-Huggins-Fumi-Tosi
(BMHFT) rigid ion potential~\cite{fumi-tosi}:
\begin{equation}
     V(r_{ij}) = \frac{Z_\alpha Z_\beta}{r_{ij}}
              + A_{\alpha\beta}\exp(-B r_{ij})  -
            \frac{C_{\alpha\beta}}{r_{ij}^6}  - \frac{D_{\alpha\beta}}{r_{ij}^8}.
\end{equation}
Here $\alpha$ and $\beta$ stand for either $+$ or $-$,
$Z_{\alpha}$ and $Z_{\beta}$ are the ionic charges ($+1$ for $Na$ and $-1$ for
$Cl$), the first term is the Coulomb interaction energy,
the second is the short-range Pauli repulsion,
and last two terms are induced dipole-dipole and dipole-quadrupole van der Waals 
interactions.The values of the parameters are reported in Table~\ref{tab}.

\begin{table}
     \begin{tabular}{l|c|c|c}
     \hline\hline
     & Na-Na & Cl-Cl & Na-Cl\\
     \hline
     A (eV)        & 424.097       & 3488.998        & 1256.31 \\
     B (\AA$^{-1}$)  & 3.1545      & 3.1545          & 3.1545   \\
     C (eV~\AA$^6$) & 1.05          & 72.5            & 7.0     \\
     D (eV~\AA$^8$)  & 0.499         & 145.427         & 8.676    \\
     \hline\hline
     \end{tabular}
     \label{tab}
     \caption{Parameters of Born-Mayer-Huggins-Fumi-Tosi potential for NaCl.}
\end{table}
%
Polarization forces~\cite{madden}, though not negligible for the quantitative
description of molecules, low-density ionic fluids, and other properties,
do not appear to be crucial in the present context. They are neglected in
order to concentrate on the BMHFT model, whose greater simplicity allows a much
more extensive computational exploration of the wetting
properties~\cite{note2}.

Bulk systems were studied at constant volume with cubic simulation
cells comprising up to 10000, but more typically 3000$\div$5000, NaCl molecular
units.
Surfaces were studied with periodically repeated slabs -- consisting of
$12 \div 24$ planes -- separated by about
$80$~\AA\ of vacuum. The long range Coulomb interaction between
ions was treated in full. We implemented a 3D Ewald summation with repeated
slabs and periodic boundary conditions along the direction $z$ orthogonal to
the surfaces as well as parallel to the $(x,y$) surface plane.
The alternative choice of a single slab with 2D Ewald summation, physically
more natural, was discarded as computationally more demanding~\cite{ewald}.
In our repeated slab 3D Ewald scheme, the
large vacuum thickness is required in order to reduce spurious electrostatic
couplings of instantaneous fluctuating dipoles in liquid surfaces facing one
another across the vacuum gap. Moreover, the Ewald sums were carried out
for better convergence with conducting (``tin foil'') boundary conditions in the
in-plane directions and with insulating or vacuum boundary conditions
in the direction normal to the surface (see Ref.~\cite{ewald}).
We performed both microcanonical and canonical simulations. In canonical
runs, temperature was controlled by velocity-rescaling.
Despite the size and time limitations imposed by long range forces, great
care was taken to run simulations long enough for a clear equilibration,
typically 100$\div$300 ps at \Tmelt, but longer when required.
Finally, in all simulations involving the solid, whether as a bulk
crystal, or at solid-liquid coexistence, or as a solid slab, we adjusted
the cell size at each temperature so as to enforce the theoretical
equilibrium lattice parameter independently computed
as the one that gave a vanishing bulk stress (see Sec.~\ref{sec:bulk_solid}).

\section{Bulk properties}

\subsection{Bulk Solid NaCl}\label{sec:bulk_solid}
\begin{figure}[!ht]
     \includegraphics[width=0.45\textwidth]{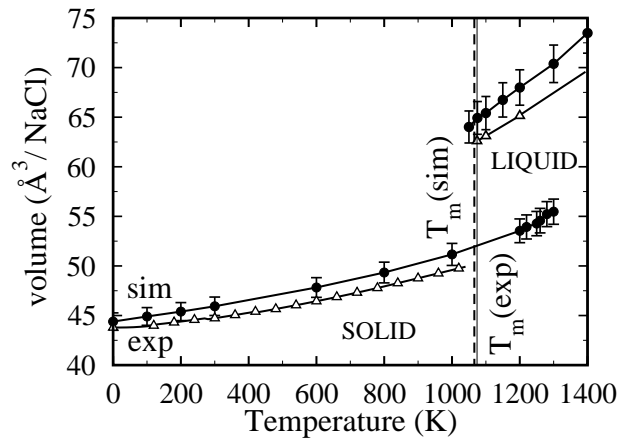}
     \caption{NaCl volume expansion vs temperature. Both the theoretical
        (filled circles) melting temperature and change of the volume at 
     \Tmelt\, are very similar to the experimental ones (empty triangles).}
     \label{volume}
\end{figure}

First of all, we verified how faithfully the bulk properties of solid
and liquid NaCl are reproduced by the BMHFT potentials.
Thermal expansion of the solid is dictated by the increase of equilibrium
lattice spacing at zero pressure as a function of temperature.
We performed several simulations at constant temperature and volume, and
computed the equilibrium lattice spacing by seeking the cell size
that yielded vanishing pressure at each given temperature.
The overall interpolated
result is shown in Fig.~\ref{volume}, where it is compared with experimental
data from various sources~\cite{landolt}. The temperature dependent cell
size obtained in this manner was subsequently enforced in all subsequent
simulations involving solid NaCl, as mentioned earlier.
The 298\,K calculated equilibrium lattice spacing and linear expansion
coefficient $\alpha = (3V)^{-1} dV/dT$ were
5.683~\AA\ and 40.5$\cdot$10$^{-6}$~K$^{-1}$ respectively
(5.635~\AA\ and 38.3$\cdot$10$^{-6}$~K$^{-1}$ are the experimental
values~\cite{landolt}).

At higher temperature, notably above 600 K, the anharmonicity is
seen to get stronger, and expansion becomes somewhat nonlinear. It
should be noted that our procedure continues to work, and is in
fact very accurate at these higher temperatures. The simulated bulk solid
remains locally stable and does not spontaneously melt (at least for our
cell sizes and for times within 200 ps) until a maximum temperature \Tspin\ 
$\sim$ 1305~K.
This maximum bulk metastability  temperature, necessarily higher than the ordinary
melting temperature \Tmelt, approximates the ``spinodal'' temperature of bulk 
NaCl, defined as
the point where the solid phase ceases to be a local free energy minimum
(for example by losing its mechanical stability). As we will see further
below, this bulk spinodal temperature is indeed predicted well above the melting
temperature, precisely \Tspin\ $\sim$ \Tmelt\ + 240~K in the BMHFT model.
It would be interesting to check this predisction by studying, e.g., spontaneous
nucleation of the liquid inside the solid, locally heated by for example
two crossed laser beams.

\begin{figure}[!ht]
     \includegraphics[width=0.4\textwidth]{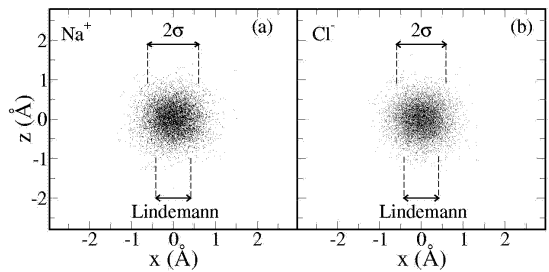}
     \caption{The instantaneous displacements of the Na$^{+}$(a) and Cl$^{-}$ (b)
     ions in simulated bulk NaCl at 1066 K (bulk melting point of the Tosi-Fumi
     model). The displacement distributions are well fit by gaussian distributions
     whose widths are indicated. The much smaller widths predicted by the
     Lindemann melting criterion
     $(\overline{\Delta r^{2}})^{1/2}\sim 0.14 \times a$,
     where $a$ is the interatomic distance, are also shown.}
     \label{rmsd}
\end{figure}

Root mean square displacements $(\overline{\Delta r^{2}})^{1/2}$ (RMSD)
of Na$^+$ and Cl$^-$ ions were extracted from
the high temperature bulk NaCl solid simulations.
As shown in Fig.~\ref{rmsd} the value at 1066~K is
0.61~\AA\ for Na and 0.59~\AA\ for Cl,
comparable with experimental estimates of 0.5~\AA\ and 0.48~\AA\ at the
melting point~\cite{lindNaCl}.
The corresponding calculated Lindemann ratios at 1066~K
\mbox{$\delta = (\overline{\Delta r^{2}})^{1/2}/a$}, ($a$ is the interatomic
distance) are 20\% and 22\% , compared with experimental estimates at \Tmelt\
obtained from Debye-Waller factors of 17\%\, and 20\%~\cite{lindNaCl}.
Simulation appears to slightly overestimate the thermal vibration
amplitudes, but we note that the uncertainties in the experimental procedure
where RMSDs were extracted seem much larger than this discrepancy. In any
case the RMSD values are very much larger than the typical values
between 10\% and 14\% of the Lindemann ratio for most solids at
\Tmelt. This large overshoot of the alkali halide bulk Lindemann ratios can be
rationalized by noting that whereas large thermal vibrations may very
effectively destabilize atoms inside its bulk solid cage when interatomic forces
are short ranged,  they will much less effectively do so when forces are long range
as is the case in strongly ionic solids. Simulations show that high
temperature NaCl is in the BMHFT model a strongly vibrating, strongly
anharmonic, and yet unusually stable solid. As we shall show later, the same
is true of the NaCl(100) surface.

\begin{figure}[!ht]
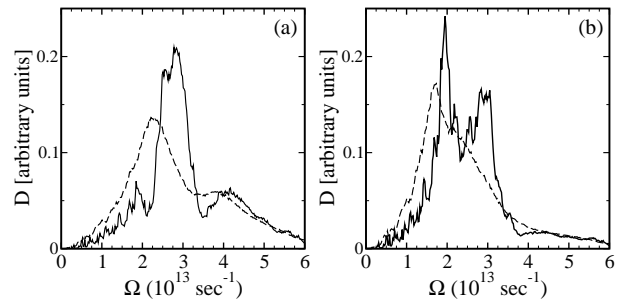

     \includegraphics[width=0.22\textwidth]{figure/statesNa.eps}
     \includegraphics[width=0.22\textwidth]{figure/statesCl.eps}
     \caption{Vibrational density of states of bulk solid NaCl for
     (a) Na$^+$ and (b) Cl$^-$ ions. Comparison between 300 ~K (solid line) and 
1000~K
     (dashed line) indicates a considerable vibrational softening.}
     \label{states}
\end{figure}

The vibrational spectral properties of warm NaCl can also be easily
extracted, for later use, by Fourier transforming the velocity-velocity
correlation functions:
\begin{equation}
     V_\mathrm{Na}(\omega) = \frac{1}{2\pi}
       \int_0^\infty dt\, e^{i\omega t}
       \left\langle \mathbf{v}_\mathrm{Na}(t) \cdot
       \mathbf{v}_\mathrm{Na}(0)\right\rangle,
\end{equation}
and similarly for Cl. The two spectral densities $V_\mathrm{Na}(\omega)$ and
$V_\mathrm{Cl}(\omega)$ extracted at 300~K and at 1066~K are shown in the
Fig.~\ref{states}.

\begin{figure}[!ht]
     \includegraphics[width=0.22\textwidth]{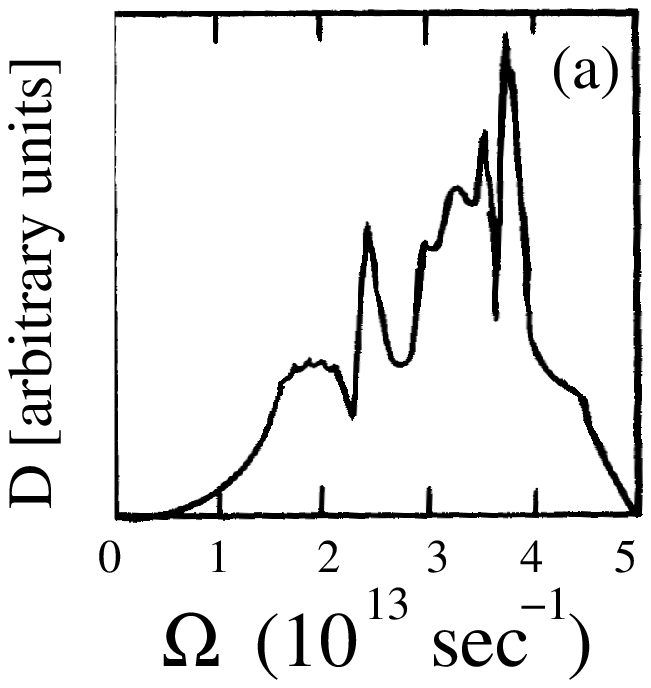}
     \includegraphics[width=0.22\textwidth]{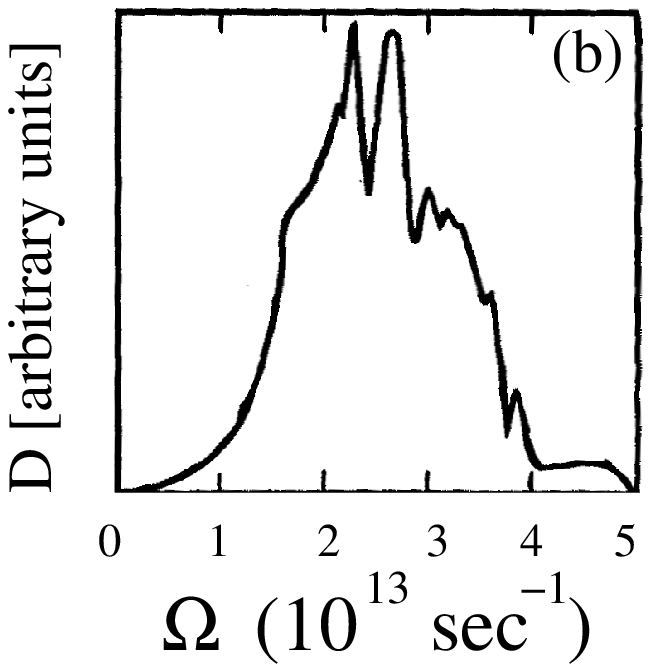}
     \caption{Vibrational density of states of solid NaCl at low temperatures
     projected on (a) Na$^+$ and (b) Cl$^-$ ions. From Ref.~\cite{hardy}}
     \label{statesexp}
\end{figure}

Generally, we note that our simulated vibrational spectra do not compare well
in the details -- even if not unreasonably in their gross features --
with the more realistic spectral densities of Ref.~\cite{hardy} taken from
the literature, and corresponding to very accurate shell model fits to
experimental phonon spectra (Fig.~\ref{statesexp}).
The rigid ion model is notoriously too crude to reproduce fine
features as the detailed vibrational spectra,  heavily affected by non
rigid ion effects~\cite{cowley}.
Nonetheless it is not unreasonable to believe
that the overall temperature evolution  of spectral densities can still be
taken as representative of the real situation. The high temperature
spectra show a considerable anharmonic softening and broadening relative to
the room temperature ones.

\subsection{Bulk Liquid NaCl}\label{sec:bulk_liquid}
\begin{figure}[!ht]
     \begin{center}
     \includegraphics[width=0.45\textwidth]{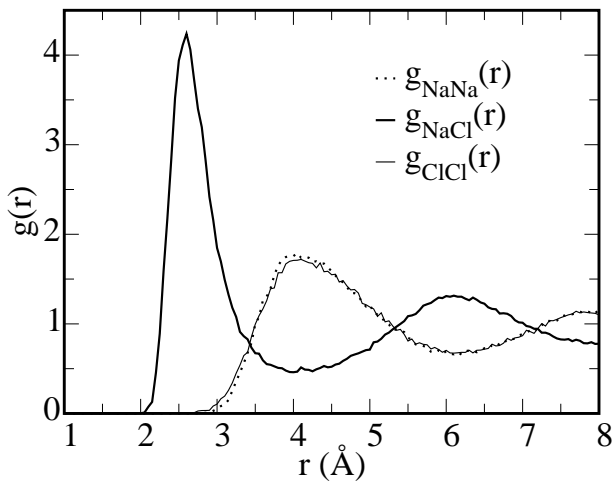}
     \caption{The partial radial distribution functions at \Tmelt= 1066 K.}
     \label{gr}
      \end{center}
\end{figure}

Simulations of bulk liquid NaCl were conducted in close analogy to those
of solid NaCl. In particular, the cell volume was adjusted isotropically
to ensure vanishing pressure at every temperature. Our overall description
of the liquid is very similar to that provided by many long-standing
studies~\cite{sangster,hansen,janz,enderby}.
The calculated liquid molecular volume is shown along with experimental
values in Fig.~\ref{volume}. The calculated volume expansion of 27\%\,
at melting compares very well with 26\%\, from experiment.
The internal structure of the liquid, notably the Na-Cl, Na-Na and Cl-Cl
radial distribution functions:
\begin{equation}
     g(r) = \rho^{-1} \left\langle\sum_{ij} \delta(\mathbf{r}-\mathbf{R}_{ij})
       \right\rangle,
\end{equation}
quantities that are long well known, are well reproduced (Fig.~\ref{gr})
by the simulation. They exhibit the charge correlation typical of molten
salts, the first peak of $g_{+-} (r)$ giving a typical Na-Cl distance in the
liquid of 2.6~\AA, about 10\% shorter than in the solid.

\begin{figure}[!ht]
     \begin{center}
     \includegraphics[width=0.45\textwidth]{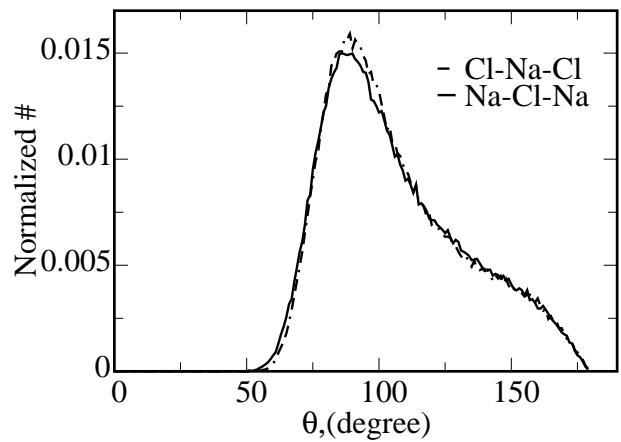}
     \caption{The angular distribution function of simulated liquid NaCl
              at \Tmelt.}\label{angular}
     \end{center}\end{figure}

   From the pair correlation functions, we may calculate the bulk liquid
coordination number:
\begin{equation}
     N_{\pm} = \rho \int_{0}^{r_m} 4\pi r^2 g_{+-} (r) dr,
\end{equation}
where the upper integration limit corresponds to the first local
$g_{+-}(r)$ minimum, $r_m = 4.0$\,\AA. We find a coordination number
\mbox{N = 4.6} in liquid NaCl at 1066~K, to be compared with the much higher
\mbox{N = 6} coordination in the defect-free bulk solid.
Experimental estimates of the liquid coordination number vary in a broad
range from \mbox{N = 4.7}~\cite{janz} to \mbox{N = 5.8}~\cite{enderby}.
According to our simulation an ion in the liquid is surrounded by a cage
consisting of only 4.6 ions of the opposite charge, albeit considerably
closer than in the solid. In order to better understand the nature of the effective
ion cage in the liquid we studied the Na-Cl-Na and Cl-Na-Cl angular
distribution in the bulk liquid:
\begin{equation}
     P(\theta) = \left\langle\sum_{ijk} \delta\left(\theta -
     \arccos\frac{\mathbf{R}_{ji}\cdot\mathbf{R}_{jk}}
                 {|\mathbf{R}_{ji}||\mathbf{R}_{jk}|}\right)
     \right\rangle,
\end{equation}
where $|\mathbf{R}_{ij}|, |\mathbf{R}_{jk}| < r_{m}$.
The result shown in Fig.~\ref{angular} exhibits a main peak at 90$^{\circ}$
and only a weaker shoulder around 150$^{\circ}$). Similar results had been
obtained by Amini~\cite{amini}. The dominant 90$^{\circ}$
peak indicates in particular that the local ionic cage surrounding an ion of
the opposite charge is, even in the liquid, still roughly cubic, or more
precisely octahedral, as in the solid.
The octahedral cage linear size is smaller by about 10\% than that
of the solid. The 4.6 surrounding ions are distributed on the 6 corners
of this cage, which therefore contains on average about 1.4 vacancies.

\noindent
The diffusion coefficient:
\begin{equation}
     D = \lim_{t \to \infty} \frac{1}{6 t} \left\langle\sum_{i}|\mathbf{R}_i(t) -
       \mathbf{R}_i(0)|^2 \right\rangle,
\end{equation}
was also computed during simulations in order to check the general quality of
description  of the liquid. We found
\mbox{D = 10.53 $\cdot$ 10$^{-5}$ cm$^2$/s} at \mbox{T = 1300~K},
which compares well with
experimental values like \mbox{D = 8.6 $\cdot$ 10$^{-5}$ cm$^2$/s}
at \mbox{T = 1121~K}~\cite{oconnel}, as well as with those from previous
MD simulations, such as \mbox{D = 9.5 $\cdot$ 10$^{-5}$ cm$^2$/s} at
\mbox{T = 1267~K}~\cite{hansen}.

Based on all the above we conclude that the BMHFT potential
description of bulk liquid NaCl is on the whole very good.

\subsection{ Bulk Melting Temperature}
The melting temperature of the BMHFT model of bulk NaCl is obtained
directly from simulations of solid-liquid coexistence at zero
pressure~\cite{extrapolated}.
At constant temperature, the interface in a bulk system which is
roughly half solid and half melted (actually there are two SL interfaces
because of periodic boundary conditions) will drift with simulation time one way
or the other so long as the system is away from the melting temperature
\Tmelt, and will only remain stationary at $T$ = \Tmelt.
At constant energy, the average temperature $\langle T\rangle$ will instead
slowly drift toward the melting temperature \Tmelt~\cite{furio,method}.

\begin{figure}[!ht]\begin{center}
    \includegraphics[width=0.32\textwidth]{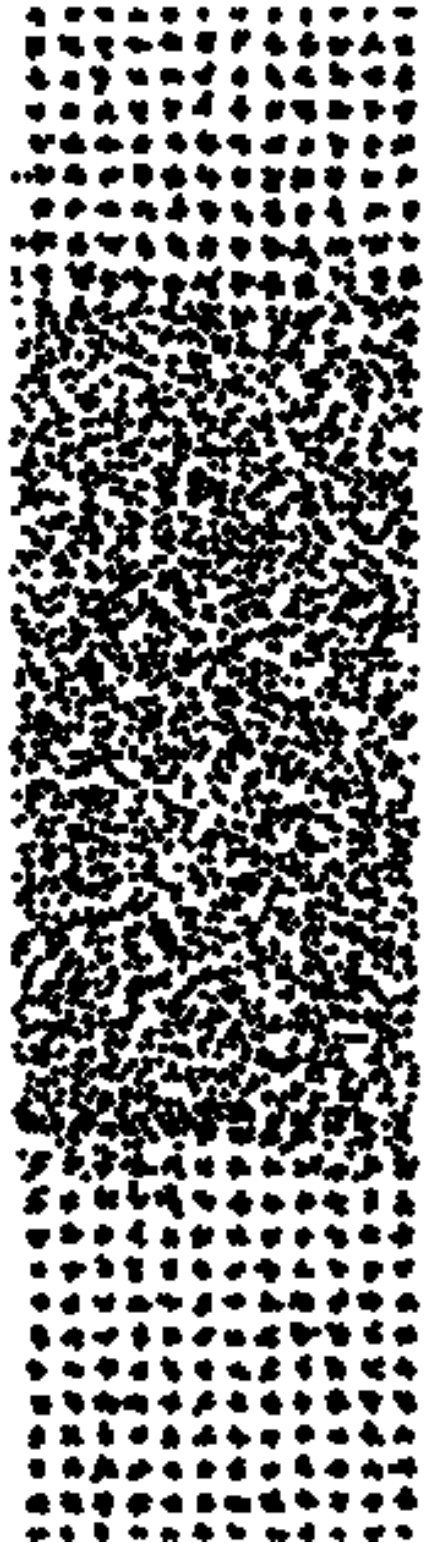}
    \caption{Simulated coexistence of bulk solid and liquid. Periodic boundary
    conditions on all sides.}
    \label{sandwich}
\end{center}\end{figure}

\begin{figure}[!ht]
     \includegraphics[width=0.25\textwidth]{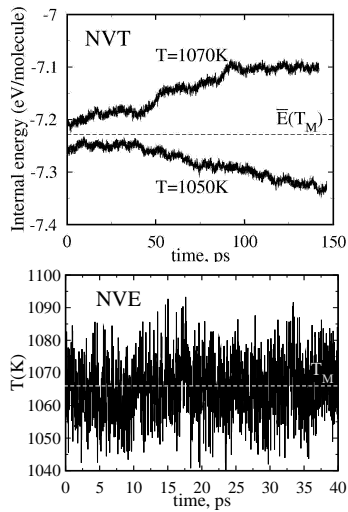}
     \hfill
     \includegraphics[width=0.25\textwidth]{figure/nve.eps}
     \caption{(a) Canonical evolution of the internal energy at liquid-solid
     coexistence. The decrease of internal energy at $T$ = 1050~K indicates that
     the SL interface moves toward recrystallization. Conversely, the increase
     of internal energy at $T$ = 1070~K indicates that the liquid phase grows
     at the expenses of the solid phase.
     The internal energy for the BMHFT melting point of bulk NaCl lies in
     between.
     (b) Final microcanonical run showing the exact location of the
     melting temperature.}
     \label{NVT}
\end{figure}

We started with a crystalline bulk made up of 2880 molecular units, a
geometry comprising 6$\times$6$\times$20 conventional 4-molecule cubic NaCl cells
along $(x,y,z)$ respectively.
Periodic boundary conditions (PBC) were assumed in all
directions, and the volume was constant during each simulation. However,
the cell size was adjusted with temperature, so as to enforce zero stress
at each temperature, as detailed below. After equilibration in
proximity of the presumed melting point (\mbox{$T \approx$ 1100 K}) about one
half of the 20 layers were melted by bringing them selectively at a higher
temperature,
while the remaining atoms in the solid phase are kept fixed on their
positions.
The liquid and the solid halves initially out of equilibrium were subsequently
let evolve to their mutual equilibrium. The in-plane cell size, and with it
the in-plane solid lattice spacing were kept fixed at their solid equilibrium
value previously established for that temperature as shown in
Fig.~\ref{volume} of Subsec.~\ref{sec:bulk_solid}.
The linear cell size in the z-direction perpendicular to the solid-liquid
interfaces was subsequently adjusted after each equilibration so as to
cancel the uniaxial stress normal to the interfaces,
compensating in particular the solid-liquid volume expansion.
When the system was finally equilibrated by means of canonical
MD runs (see Fig.~\ref{sandwich}), we observed the anticipated drift of the
solid-liquid interfaces in opposite directions at 1050~K and
1070~K, indicating that the melting temperature must fall in between
(Fig.~\ref{NVT}a).

To sharpen up our estimate of \Tmelt\ we then carried out delicately
selected microcanonical simulation runs between $E = -$7.2 and $-$7.25
eV/molecule. Our final result was \mbox{\Tmelt = 1066 $\pm$ 20~K}
(Fig.~\ref{NVT}b).
The conservatively large error bar quoted is mainly due to a residual
pressure uncertainty of $\sim$0.5 kbar. 
The real uncertainty is probably smaller, since we also found later
that systems with free surfaces, and thus with a pressure more accurately
close to zero, melt in bulk within 5~K of 1066~K. We also estimate that
the error of \Tmelt\ due to other factors, including small system size,
and fluctuations, to be smaller than 20~K.
For instance, the melting temperature of an unrealistically small system
made of 6$\times$6$\times$4 molecules was found to be $\sim$ 1050~K.
Finally our value of the melting temperature, obtained in a totally unbiased
manner~\cite{droplet}, is in essentially perfect agreement with
\mbox{\Tmelt\ = 1064 $\pm$ 14~K}
independently obtained by thermodynamic integration by Frenkel's
group~\cite{frenkel}.

\clearpage

The latent heat (enthalpy of melting) is estimated as
the change of the internal energy at \Tmelt\ and found to be 0.29
eV/molecule.
The calculated entropy jump at melting is thus
$\Delta S_{m} = L/\Tmelt =$ 6.32 $\pm$ 0.1 $k_{B}$/molecule.
A slope of the melting line (Clausius-Clapeyron relation)
$    \frac{dP}{dT} = \frac{L}{\Tmelt(v_{l}-v_{s})}$
of 0.0311 kbar/K is then obtained. All these calculated results are in quite
good agreement with experimental data. Tab.~\ref{tab_bulk} summarizes the
calculated values and their comparison with experiment.

\begin{table}\begin{center}
    \begin{tabular}{l|c|c}
    \hline\hline
    & Simulation & Experiment \\
    \hline
    \Tmelt (K) & 1066$\pm$20 & 1074 \\
    $\Delta V$ & 27\% & 26\% \\
    L (eV/molecule) & 0.29 & 0.29 \\
    $\Delta S_m$ ($k_B$) & 6.32 & 6.38 \\
    dP/dT (kbar/K) & 0.0311 & 0.0357~\cite{dpdt} \\
    \hline
    $\alpha$ (10$^{-6}$ K$^{-1}$) & 40.5 & 38.3 \\
    RMSD (\AA) & 0.60 & 0.49 \\
    $\delta$ & 20--24\% & 17--20\%~\cite{lindNaCl} \\
    \hline\hline
    \end{tabular}
    \caption{High temperature properties of NaCl.
    \Tmelt\ is the melting temperature; $\Delta V$ is the volume jump at the
    melting point; $L$ is the latent heat of melting; $\Delta S_m$ is the entropy
    variation at the melting point;
    $dP/dT$ is the resulting Clausius-Clapeyron ratio
    at the melting point. $\alpha$ is the linear thermal expansion coefficient;
    RMSD is the root mean square displacement of atoms in the bulk solid at the
    melting point; $\delta$ is the RMSD over the Na--Cl distance, for the
    Lindemann melting criterion.}
    \label{tab_bulk}
\end{center}\end{table}

\section{Crystalline NaCl(100): a Non-melting Anharmonic Surface}
\label{sec:crystalline_100}
\begin{figure}[!ht]
     \begin{center}  
     \includegraphics[width=0.45\textwidth]{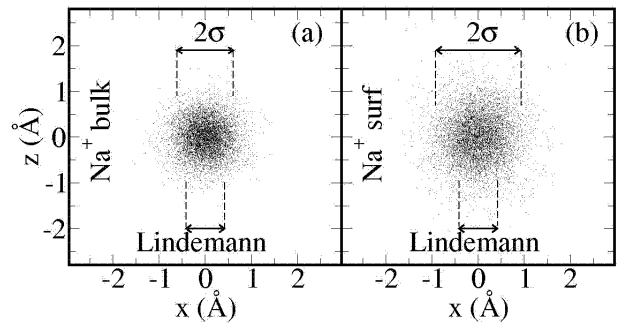}
     \caption{Solid NaCl instantaneous displacements of Na$^+$ ions in (a) bulk
     and (b) first surface layer at \Tmelt.}
     \label{msdNa}
     \end{center}
\end{figure}

\begin{figure}\begin{center}
    \includegraphics[width=0.45\textwidth]{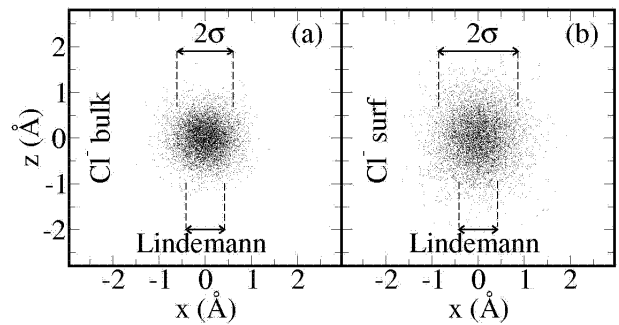}
    \caption{Solid NaCl instantaneous displacements of Cl ions in (a) bulk
    and (b) first surface layer at \Tmelt.}
    \label{msdCl}
\end{center}\end{figure}

Satisfied with the above description obtained for bulk NaCl, we moved on to
study the NaCl surfaces, by simulating slabs as described earlier.
In slab simulations, we found the defect free NaCl(100) to warm up
uneventfully, and to remain solid and totally dry up to \Tmelt.
The root mean square displacements (RMSD) of the first layer Na$^+$ and
Cl$^-$ ions
at \Tmelt\ were extracted and are shown in Fig.~\ref{msdNa} and
Fig.~\ref{msdCl}.

Unsurprisingly, the surface ions vibrate more than bulk ions. The ratio of
surface/bulk RMSD at \Tmelt\ is here 1.5. For comparison, in the
Pietronero-Tosatti model~\cite{pietronero} the critical surface melting
value of this ratio is $\sim$ 1.6. Moreover, vibrations are somewhat more extended
in direction perpendicular to the surface than in the in-plane direction.

\begin{figure}[!ht]
     \includegraphics[width=0.45\textwidth]{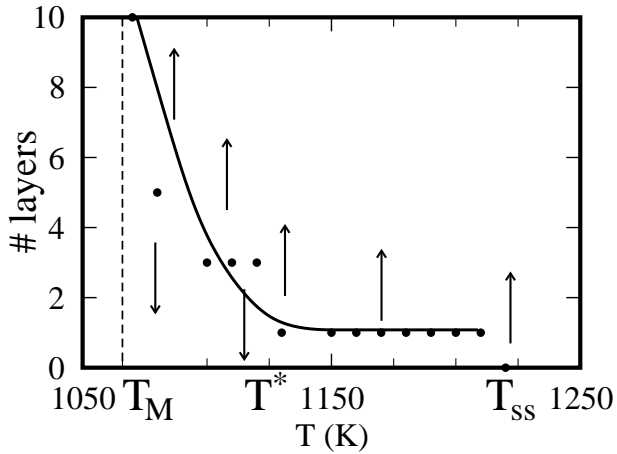}
     \caption{Critical liquid film thickness vs. temperature for metastable
     solid NaCl(100) above \Tmelt. This corresponds to the effective thickness
     of the free energy barrier for the nucleation of the NaCl melt at the solid
     surface.}
     \label{nucl}
\end{figure}

We subsequently verified that even \emph{above} \Tmelt\ the NaCl(100) surface
remained crystalline in a metastable state for at least 200 ps.
In this overheated surface regime, solid NaCl(100) was found to possess a
thick nucleation barrier against melting up to about \Tstar\ = 1115~K $\sim$
\Tmelt\ + 50~K, in the following sense. When a thin surface film consisting
of $l$ atomic layers was artificially melted, and then let evolve to
equilibrium at a grid of temperatures $T$ $\geq$ \Tmelt, the melted film
was seen to spontaneously recrystallize for $l \leq l_{crit}(T)$, with 
$l_{crit}(T) \geq 1$ for
$T \leq$ \Tstar. (See Fig.\ref{nucl}). That indicates that in this temperature range the
overheated solid slab is locally stable, with a free energy barrier
against overall melting~\cite{carnevali}.

Above \Tstar, overheating of crystalline NaCl(100) persisted
until a higher ``surface spinodal temperature'' \Tss
$\simeq$ 1215~K $\simeq$
\Tmelt + 150~K, now however with only a thin nucleation barrier
$l_{crit}(T)$ of a monolayer.

Only at \Tss, as high as 150~K above \Tmelt, does solid NaCl(100) become
locally unstable and melt spontaneously. We find the pronounced metastability
of this solid surface to persist at least up to \Tstar, or $\sim$ 50~K above
\Tmelt\ even in presence of common surface defects, such as molecular
vacancies or steps.

We have thus characterized NaCl(100) as a clear case of surface non-melting.
This is a prediction that deserves to be tested in experiment. For a short
enough time, it should be possible to raise the temperature of NaCl(100) by
at least 50~K above \Tmelt, without any liquid spontaneously nucleating at
the surface. Of course, the solid surface will sublime very strongly at
these temperatures. The experimental equilibrium vapor pressure of NaCl at
\Tmelt\ is 0.345 mmHg, a large value. As a consequence, surface 
vacancies will form and surface steps will flow due to evaporation (and to
re-condensation when working at solid-vapor equilibrium).
We note that the time scale for this kind of surface evolution is a much
slower one than that addressed here.
The experimental rate of evaporation found from empirical expression~\cite{evap}
valid for a variety of materials is about
4$\cdot 10^{-5}$ gr/cm$^2$ s.
For our typical 10$\times$10 sized surface (200 surface NaCl units)
of area around 3500 \AA$^2$, and using an experimental vapor pressure of
0.345 mmHg the evaporation rate is $\sim$ 1.7 $\cdot 10^{5}$ s$^{-1}$. This indicates a typical
evaporation time many orders of magnitude larger than our typical simulation
time.

That explains why in practice we never even observed in simulation a
spontaneous evaporation event off the solid surface. This however does not invalidate the
significance of the simulation results. At any given instant of
a realistically long time evolution of NaCl(100), there will be of course
molecular evaporation and step flow, but that will still leave defect
free terraces much larger on average than those we simulated.
These terraces will be fully crystalline even well above \Tmelt, displaying
precisely the microscopic non-melting behavior described above.

These considerations however suggest that surface non-melting in alkali halides
could not easily be pursued with static or nearly static experimental probes,
because sublimation and condensation will influence and possibly spoil the
outcome. Perhaps the fast laser tools already employed for metals~\cite{ali}
could be brought to bear on this case too. The high vapor pressure also
suggests techniques that do not rely on ultra-high vacuum~\cite{metois}.
We are currently considering in addition hard tip sliding friction as a
tool, with results to be described in forthcoming paper.

\section{Solid Surface Free Energy: Thermodynamic Integration}
In order to prepare for our thermodynamical assessment of the wetting of
NaCl(100) by liquid NaCl, we eventually need to know according to 
eq.~(\ref{eq:young1})
the solid-vapor interface free energy
$\gammaSV$ at the melting point.

\begin{figure}\begin{center}
    \includegraphics[width=0.45\textwidth]{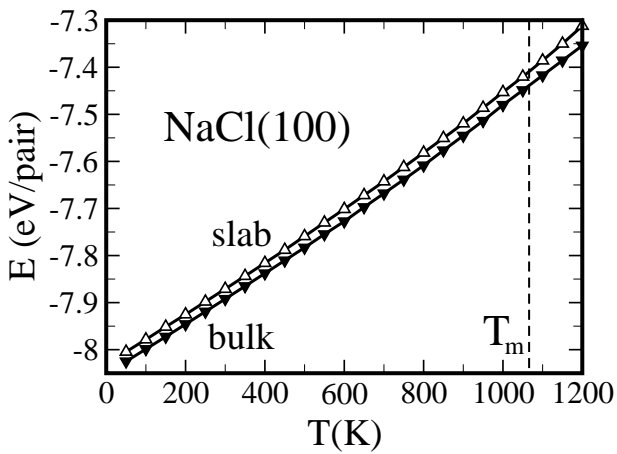}
    \caption{Bulk (filled triangles) and slab (empty triangles) internal energies
    as a function of temperature. The slab consisted of 1440 NaCl molecules.}
    \label{int1}
\end{center}\end{figure}

We calculated $\gammaSV$ through thermodynamic
integration~\cite{frenkel96} using the following relation:
\begin{equation}\label{eq:integration}
      \left( \frac{\partial(F/T)}{\partial(1/T)} \right)_{N, V} = E,
\end{equation}
where $F$ is the free energy and $E$ the internal energy.
We simulated for this purpose a 2880 molecule bulk system and independently
a slab comprising 1440 molecules and an equivalent volume of vacuum.
We kept in this manner the all-important Ewald sum convergence unchanged,
(we judge the implied extra error due to size effects to be negligible
by comparison). By integrating the internal energy over $(1/T)$ up to
temperature T we separately obtained the bulk and the slab free energies per
molecule up to \Tspin\ and to \Tss\ respectively.
We did not explicitly include quantum freezing effects at low temperatures,
because they represent a small correction in comparison with large thermal
effects at the melting point.
Nevertheless since quantum freezing takes place at temperatures
$T \lesssim (1/4)\theta_{D}$, ($\theta_D$ is the Debye temperature), we started
integration at $T_i = (1/4)\theta_{D}$ which is $\simeq$ 50~K for NaCl,
therefore using the $T$ = 50~K state as a reference.
\begin{equation}
     \frac{F(T)}{T} - \frac{F(T_i)}{T_i} =
     \int_{1/T_i}^{1/T} E(T')\, d\left(\frac{1}{T'}\right).
\end{equation}
The bulk and slab internal energies as a function of temperature, are
shown in Fig.~\ref{int1}. After integration, the difference between
slab and bulk free energies per unit surface
area (accounting the presence of two surfaces) equals the surface energy.

\begin{figure}\begin{center}
    \includegraphics[width=0.45\textwidth]{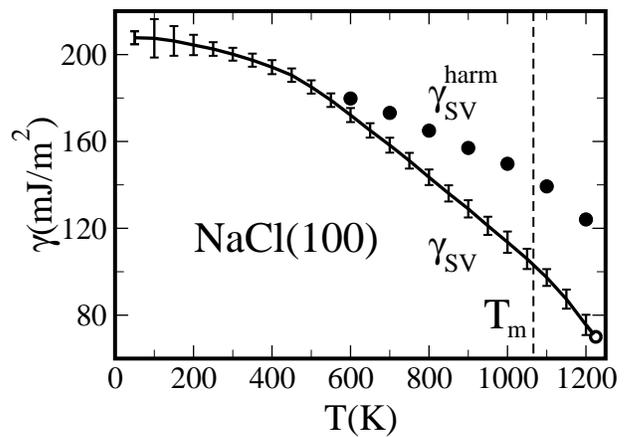}
    \caption{The solid surface free energy from thermodynamic integration
    (line) and from the effective harmonic approximation (dots).}
    \label{int2}
\end{center}\end{figure}

The NaCl(100) surface free energy calculated in this manner is displayed in
Fig.~\ref{int2}. Starting at low temperature, we note from the start a low
surface energy, reflecting the excellent charge order of this
surface. Upon increasing temperature, there is an
important thermal drop of the surface free energy, especially fast for
\mbox{$T \gtrsim$ 600~K}, indicating considerable additional anharmonicity
of the
solid surface relative to bulk solid NaCl. A good question is what part of
that anharmonicity can be ascribed to temperature-dependent effectively
harmonic vibrations, and what cannot.

To judge on that, we computed separately the bulk and slab vibrational spectra as a
function of temperature. Using simulation trajectories in a bulk and
in a slab comprising exactly the same number of molecules, we extracted the
Fourier transformed velocity autocorrelation functions of ions 
$V_\mathrm{Na}(\omega)$
and $V_\mathrm{Cl}(\omega)$.  The vibrational spectra are obtained as:
\begin{equation}
    A(\omega) = \frac{m_\mathrm{Na}}{k_{B} T} V_\mathrm{Na}(\omega) + 
\frac{m_\mathrm{Cl}}{k_{B} T}
      V_\mathrm{Cl}(\omega).
\end{equation}
By treating both the bulk and the slab spectra as a collection of
harmonic oscillators, we obtained an effective
surface vibrational free energy by subtracting their respective harmonic
free energies off one another, and dividing the outcome by two, for two
surfaces.

For increasing temperatures, the surface component of the slab spectra
displays a stronger anharmonic softening than the bulk. This as anticipated
gives rise to an ``effective harmonic'' drop of surface free energy, as shown
by dots in Fig.~\ref{int2}. We conclude that whereas about half the total
anharmonic free energy decrease can be ascribed to the effective surface
vibrational free energy and in particular to the surface frequency softening
with temperature, the remaining half cannot be accounted for in this way,
representing ``hard'' anharmonicity.

In conclusion the surface free energy of NaCl(100) approximately halves
its value from $\sim$~206~mJ/m$^2$ at 50~K to $\sim$~100~mJ/m$^2$ at \Tmelt. Such 
a large decrease of the already unusually small low-$T$ surface energy results in an
exceptionally stable solid surface. While the physical reason for a low
surface energy at low-$T$ is clearly the perfect charge ordering, that
for its large thermal decrease is the ability of rocksalt and of
its surface to sustain exceptionally large anharmonic vibrations without
loss of mechanical stability.

\section{The Liquid NaCl Surface}

We proceeded next to study the liquid NaCl surface, or more correctly
the liquid-vapor interface. There are earlier simulations of this liquid
surface in the BMHFT model, notably by Heyes~\cite{heyes}, and there is
theory work in the Restrictive Primitive Model~\cite{groh,evans}. The more
sophisticated recent studies of Aguado et al.~\cite{madden} of the liquid KI
surface emphasize the role of polarization forces, and also summarize
earlier work.

Notwithstanding that, our scope here is to pursue a homogeneous comprehensive
study of all surface properties in a single simple model, and we therefore
carried out a fresh study of NaCl in the BMHFT model, now with a larger size
scale than that of Heyes~\cite{heyes}.

Starting with the same 6$\times$6$\times$12 solid slab used in the previous section for the study
of NaCl(100), temperature was raised above the surface spinodal temperature
causing the slab to melt. Subsequently, the liquid slab was gradually cooled down to
\Tmelt, and equilibrated for about 50~ps, after which correlations were
examined.

Here too, evaporation of molecules and molecular dimers was very seldom observed,
as an extremely rare event in our liquid surface simulations. The
occasionally evaporated molecule traveled in vacuum to recondense within
a relatively short time, but had no chance to get otherwise 
equilibrated during the flight.
The occasional molecular evaporation or condensation events were very quickly
``forgotten'' in the chaotic liquid surface dynamics. In this regime
they therefore do not appear to affect at all the overall liquid surface behavior.
This allows us to neglect our lack of an evaporation/condensation
statistics, in that its inclusion would not alter the liquid surface properties 
to be
extracted by the simulation.

\begin{figure}[!ht]
     \includegraphics[width=0.45\textwidth]{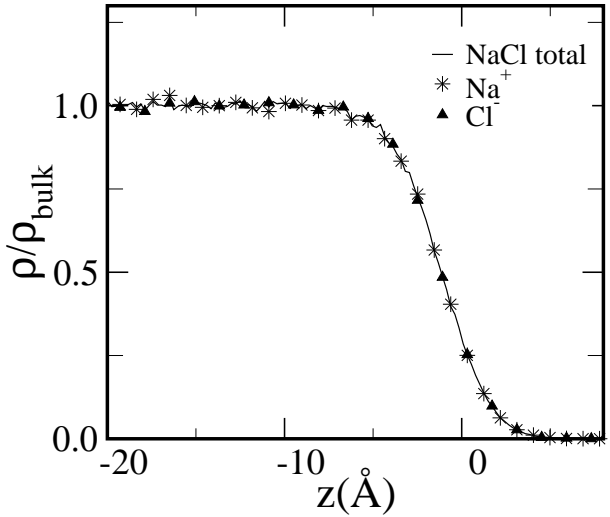}
     \caption{Na, Cl and total density profiles of the simulated liquid
     surface of NaCl at $T=\Tmelt$. Layering oscillations and surface dipoles
     are absent. Capillary fluctuations are very modest for our small cell
     size, and the large width of the liquid-vapor interface is a genuine
     local property of the surface.}
     \label{profile}
\end{figure}

Because of the strong charge correlations in the bulk liquid, one might
naively but not unreasonably have expected this liquid surface to be
structured, maybe layered as in the metals~\cite{physrep}, perhaps
displaying a surface dipole~\cite{evans}.
The actual liquid local surface density profile $\rho_{+-}(z)$ obtained for both
ionic species is shown in Fig.~\ref{profile}. All profiles are remarkably
coincident and smooth, thus -- as one could also see from earlier MD
studies~\cite{heyes, madden}-- totally devoid of layering. Moreover the Na and Cl
profiles are superposable with very great accuracy, thus the liquid surface
displays no static average dipole either, whereas the local time and space
dependent dipole fluctuations are large.

We worried that the large apparent surface spatial width should really
represent the local liquid surface structure, and not, {\it e.g.,}
simply reflects long wavelength capillary fluctuations. For that purpose we
carried out additional liquid slab simulations with alternatively a much
smaller cell of size 4$\times$4$\times$4, or a much larger one of size
10$\times$10$\times$15. We found that the resulting surface density
profiles remain essentially the same. We conclude that the additional capillary 
broadening
of the surface profile will only show up (logarithmically) for much larger 
sizes, and that the
observed surface diffuseness is indeed intrinsic.

\begin{figure}[!ht]\begin{center}
    \includegraphics[width=0.45\textwidth]{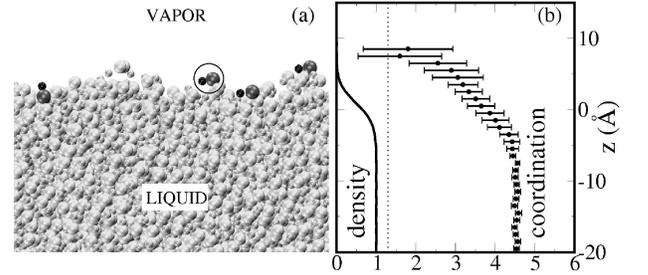}
    \caption{a) Simulation snapshot of the NaCl liquid surface at \Tmelt.
    Note the large thermal fluctuations, with some nearly molecular
    configurations highlighted in the outermost region; b) Coordination number
    $N(Z)$ and density profile, confirming a very smooth crossover from
    liquid ($N$=4.6) to molecular vapor ($N$=1.3 dotted line).}
    \label{liq}
\end{center}\end{figure}

The nature of diffuseness of the NaCl liquid surface is clarified by the
simulation snapshot of Fig.~\ref{liq}a, showing very pronounced local
thermal fluctuations in the instantaneous surface profile. This picture,
suggestive of a low surface tension, high entropy surface, is in apparent
contradiction with the massive non-wetting of solid NaCl(100) by its own
melt, the latter implying a relatively high liquid surface tension.

In order to clarify the situation, we undertook a direct calculation of the
surface free energy $\gammaLV$, equal to the surface tension, of liquid NaCl. The
calculation was done by evaluating surface stress of the slab (which has two
equivalent surfaces) via the Kirkwood-Buff formula:
\begin{eqnarray}\label{KBnew}
\gammaLV &=&
\frac{1}{2}\int_{0}^{L_{z}} dZ[\sigma_{\parallel}(Z) - \sigma_{\perp}(Z)] {}
\nonumber \\
&=& -\frac{1}{8}\int_{-\infty}^{\infty} dZ \int
d^{3}\textbf{r}_{ij} \sum_{\alpha,\beta} 
\frac{x^{2}_{ij}+y^{2}_{ij}- 2z^2_{ij}}{r_{ij}} \big[\delta_{\alpha \beta} + {}
\nonumber \\
&+& \lambda(1 - \delta_{\alpha \beta})\big]
f_{\alpha\beta}(r_{ij}) g_{\alpha \beta}^{(2)} (\textbf{r}_{ij};
Z)\rho_{\alpha}(Z)\rho_{\beta}(Z){} \nonumber \\
&=&-\frac{1}{8L_{x}L_{y}}\langle \sum_{i\alpha,j\beta}
\frac{x^{2}_{ij}+y^{2}_{ij}-2z^2_{ij}}{r_{ij}}
\big[\delta_{\alpha \beta} + {} \nonumber \\
&+& \lambda (1-\delta_{\alpha \beta})\big] f_{\alpha\beta}(r_{ij}) \rangle
\end{eqnarray}
where: $(\alpha,\beta) = (+,-)$, $i\alpha$ and $j\beta$ denote ions at site $i$ or $j$, $Z$ is the distance normal to the interface,
$L_{x}, L_{y}$ are the $(x,y)$ dimensions of the supercell
and $\sigma_{\parallel}=\frac{1}{2}(\sigma_{xx}+\sigma_{yy})$ and
$\sigma_{\perp}=\sigma_{zz}$
are the tangential and normal components of the stress tensor respectively.
Here $\langle \ \rangle$ denotes a canonical average and $\sum_{i,j}$ is over
all pairs of particles. Moreover ${\textbf{r}}_{ij}= (x_{ij}, y_{ij}, z_{ij}$)
is the interatomic distance, $f_{\alpha \beta}(r_{ij})$ is the force between
atoms $i$ and $j$, $g_{\alpha \beta} ({\textbf{r}}_{ij}$; Z) are the Na-Cl, Na-Na, Cl-Cl
pair correlation function measured in a slice centered at $Z$,
$\rho_{\alpha}(Z)$ the average density of ion $\alpha$ near $Z$ and finally
$\lambda$ is a parameter here equal to one, but inserted for later use.

The calculated liquid surface tension $\gammaLV$ is shown as a function of
temperature in Fig.~\ref{stress}.
The value at \Tmelt\ is 104 $\pm$ 8 mJ/m$^2$,
in fairly good agreement with the experimental surface tension of
116 mJ/m$^2$ -- and by chance essentially identical to that of the solid.
A very large anharmonicity was shown earlier to explain the relatively
low surface free energy of solid NaCl(100). The physical
reasons that make the liquid surface tension so relatively high will be
addressed below.
\section{Liquid Surface Entropy Deficit:
            Incipient Molecular Order in the Liquid Surface}\label{sec:liqsurf}
We wish to understand the reasons for the relatively high surface free energy of
liquid NaCl. A clue is provided by a comparison of solid and liquid
surface excess \emph{entropies}:
\begin{equation}
     S_\mathrm{surf} = - d\gamma/dT.
\end {equation}
Generally, one would expect that the much looser structure and greater
freedom of ionic motion at a liquid surface should yield a larger liquid
surface entropy than that of the solid surface. Strikingly, in NaCl the
smaller calculated temperature dependence
of surface free energies shows just the reverse. We find a factor 2.6
lower surface entropy $S_\mathrm{LV}$ compared with $S_\mathrm{SV}$
of the solid surface. Let us focus on this inverted result, which
indicates  qualitatively speaking a liquid surface entropy deficit (SED).
An entropy deficit is suggestive of some form of underlying
surface short range order. The order is clearly not layering: so what is it instead?

The answer we found is that charge order, already very important in
bulk, plays a newer and enhanced role at the molecular liquid surface.
If surface thermal fluctuations are indeed very large, they are also
revealingly {\em correlated}. For a Na$^+$ ion that instantaneously moves
e.g., out of the surface, there is at least one accompanying Cl$^-$,
also moving out; and vice versa. So while the large surface fluctuations
smear the average liquid vapor density profile, bridging gently
between the liquid and essentially zero in the vapor, (Fig.~\ref{profile})
the two-body correlations, described e.g. by the the Na-Cl pair correlation
function $g_{+-}(\textbf{r})$, and by its integral, the ion coordination
number $N$, drop from values typical of the bulk liquid at \Tmelt\ to the
{\em nonzero} value of the molecular vapor. For a quantitative characterization,
we calculated a locally defined charge coordination number:
\begin{equation}
     N_{\pm}(Z) = \frac{1}{2 \delta z}\int_{Z - \delta z}^{Z + \delta z}
      \left\{ dZ' \rho(Z')
       \int_{r<r_{m}} d^{3}\textbf{r}\ g_{+-} (\textbf{r};Z') \right\}
\end{equation}
where $r_m =$ 4.0\AA\ corresponds to the first local minimum of
$g_{+-}(\textbf{r})$, and $\delta z$ is a small interval. Starting with $Z$
inside the liquid slab, where the environment is bulk-like, we recovered
$N_{\pm_L}=4.6$ \AA\ at \Tmelt, as in the bulk liquid.
Moving $Z$ across the liquid-vapor interface we found $N_{\pm}(Z)$ to drop
continuously from 4.6 downward 
(Fig.~\ref{liq}b).


In NaCl we know (even if simulation statistics is non-existing in the vapor)
that $N_{\pm}(Z)$ for large $Z$ is bounded below not by zero but by $N_{\pm_V}$,
the average value for the NaCl vapor at \Tmelt. Experimentally the NaCl
vapor consists for 69\% of molecules (N=1), 31\% of dimers, (N=2) and a
trace of trimers~\cite{vapor}.
The corresponding vapor average is $N_{\pm_V} \simeq 1.3$. Here emerges the
crucial difference between the molecular NaCl vapor and {\it e.g.,}
the atomic LJ vapor, where $N_{\pm_V}\simeq 0$. The larger the coordination
number of ions in the surface region, the less their configurational
entropy, the higher the liquid surface tension.
Hence incipient molecular order~\cite{note} could provide the reason for the SED found
for liquid NaCl.

For a test of this idea, we repeated the same Kirkwood-Buff
calculation of the surface tension done previously, now however by
slightly and artificially altering in
Eq.~(\ref{KBnew}) the value of correlations $g_{+-}$ at the
surface. Specifically, we artificially reduced to zero the weight $\lambda$
attributed to forces acting among Na$^+$ and Cl$^-$
ions for that (extremely small) fraction of outermost surface atoms whose 
coordination
number $N_\pm \lesssim N_{\pm_V} \simeq 1.3$, the mean vapor value.
The contribution of these configurations to the pressure should provide a good
measure of the the influence of incipient molecular charge ordering to
the surface free energy, in particular to the surface entropy.
Through this highly artificial but in our view illuminating
procedure, the surface internal energy (a mechanical variable) remains untouched,
and thus only surface entropy is affected.

\begin{figure}[!ht]
     \includegraphics[width=0.45\textwidth]{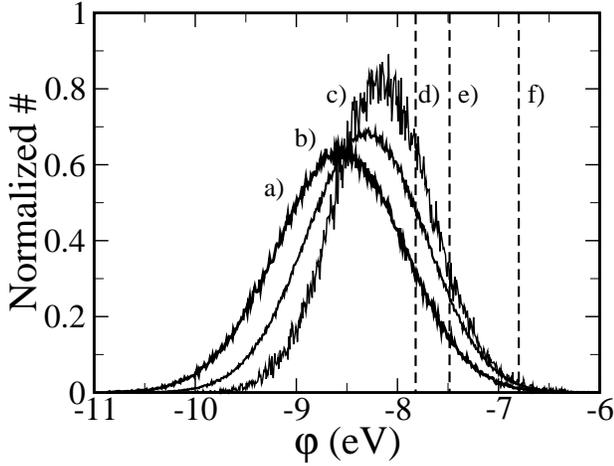}
     \caption{The electrostatic potential distribution at Na$^+$ ions in (a)
     bulk liquid NaCl; (b) solid bulk, and (c) slab liquid of thickness 70~\AA,
     at \Tmelt. Comparison the potential at with BMHFT (d) trimers
     (e) dimers and (f) monomers is also provided. Precise values of
     the average $\phi$ are
     (a) $-$8.57 eV, (b) $-$8.32 eV, (c) $-$8.21 (d) $-$7.82 eV,
     (e) $-$7.485 eV, (f) $-$6.8 eV.
     The electrostatic distributions of Cl$^-$ ions are just specular, that is
     identical to those of Na$^+$, with a plus sign.}
     \label{num}
\end{figure}

We first identified the \emph{surface} Na and Cl atoms in the simulation by means
of a simple algorithm. All ions are binned according to increasing $Z$ and are
represented by a sphere of finite radius (1.1 \AA\ in our case).
An ion is considered a surface ion when the projection on the $(x,y)$ plane
of its representative sphere is non overlapping with that of any other ion
at larger $z$. For each so identified surface ion {i}, we extracted from
the simulation the instantaneous electrostatic potential value $V_i$, a
quantity related to the coordination number, but more convenient to
calculate and to handle. As shown in
Fig.~\ref{num} the overall electrostatic potential distribution of, say, Na
ions in the liquid slab is shifted toward the electrostatic
potential value typical of the NaCl molecule, and away from that of the
bulk liquid. This shift is evidently caused by the lower coordination of
ions on the two surfaces of the slab, for the slab interior is
bulk-like. The surface Na$^+$ ion potentials for example are shifted by
$\sim 0.1$ eV on average relative to their bulk counterpart. The
electrostatic potentials of Cl$^-$ ions behave specularly, and are thus shifted
by $\sim -0.1$ eV relative to their bulk counterparts.

\begin{figure}[!ht]
     \includegraphics[width=0.45\textwidth]{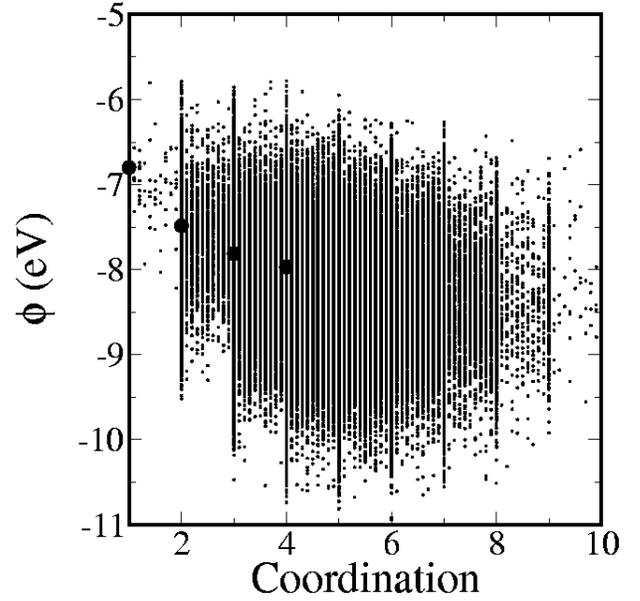}
      \caption{Electrostatic potential distribution of surface Na$^+$ ions,
      plotted versus their own instantaneous coordination number.
      Superposed is the molecular correlation between potential and
      coordination(black dots).}
      \label{monster}
\end{figure}

As the next step we established the necessary connection between average
electrostatic potentials (easily extracted from simulations, and at
least in principle easily measurable) and coordination numbers (hard to
extract from simulations,and probably harder to measure). The potential should
vary monotonically with coordination, e.g., linearly for a fixed
interatomic distance. Since in reality the Na-Cl distance increases with
coordination, the overall dependence is somewhat less than linear as shown in
Fig.~\ref{num}. A raw histogram of electrostatic potentials against
coordination numbers for the surface ions in the simulated liquid slab
at \Tmelt\ is shown in Fig.~\ref{monster}.

Finally, we computed a modified liquid surface tension by cutting off
in the Kirkwood-Buff average in Eq.~(\ref{KBnew}) the Coulomb part of the
Na-Cl force contribution of surface Na ions, through a parameter $\lambda$
of the form $\lambda = \theta(V_{0}-V_{i})$ where $\theta$ is the step
function and $V_0$  is a cutoff potential value, selecting the type
of correlations to be removed. For example,  $V_0 = -$6.8 eV cuts off up
to monomer correlations, $V_0 = -$7.485 eV cuts off up to dimer correlations,
etc. In particular we choose $V_0 = -$6.99 eV, the value that cancels correlations
for Na ions with $N \leq 1.3$, the vapor average.

\begin{figure}[!ht]
     \includegraphics[width=0.45\textwidth]{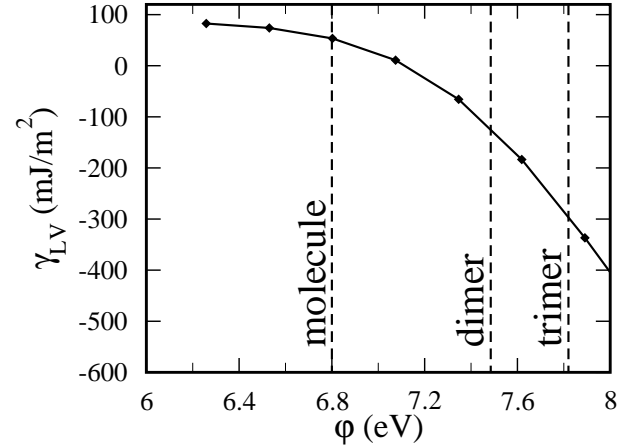}
     \caption{The decrease of fictitious liquid surface tension
     $\gamma^{*}_\mathrm{LV}$ at \Tmelt\ with increasing surface correlation
     coordination cutoff.}
     \label{liqten}
\end{figure}

\begin{figure}\begin{center}
    \includegraphics[width=0.5\textwidth]{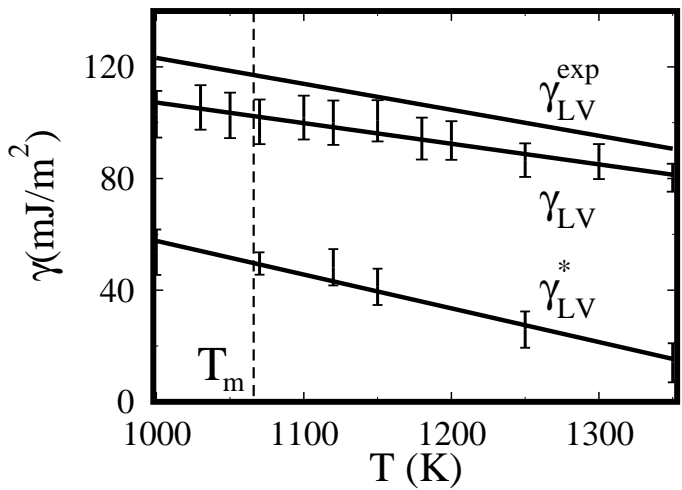}
    \caption{Liquid NaCl surface tension. Note the good agreement of the
    calculated $\gammaLV$ with experiment $\gammaLV^{exp}$. $\gammaLV^{*}$:
    artificial surface tension calculated by setting $ \lambda$ = 0 for outer
    surface atoms with coordination below 1.3 (highlighted in Fig.~\ref{liq}a).
    Once surface molecular order is removed in this way, surface entropy rises,
    surface tension drops, as shown. Solid NaCl(100) would be completely
    wet by this artificial liquid.}
    \label{stress}
\end{center}\end{figure}

With this tool, we are now able to examine more quantitatively the surface
tension contribution due to the incipient surface molecular correlations
causing \mbox{$N_{\pm}(Z) \to\, N_{\pm_V} \simeq 1.3$}, by correspondingly choosing
the cutoff potential $V_0 = -$6.99 eV.
This modification generally affects an exceedingly small fraction of
surface ions. In particular, the Na$^-$ ions affected are quite few, as
highlighted in Fig.~\ref{liq}a.
Nevertheless the partial removal operated of the surface tension contribution
due to this molecular part of surface Coulomb correlations yields a very
considerable overall surface-tension decrease, with a large drop
from $\gammaLV = 104$~mJ/m$^2$ to $\gamma^{*}_\mathrm{LV}=$ 53 mJ/m$^2$,
as shown in Fig.~\ref{liqten}, and Fig.~\ref{stress}.
This we interpret
as a direct confirmation that incipient molecular surface correlations are
indeed responsible for the liquid SED and for the resulting high surface
tension. Remarkably, since now $\gamma^{*}_\mathrm{LV}+ \gammaSL < \gammaSV$,
the surface tension drop following the hypothetical removal of short range
surface molecular correlations would actually suffice to
drive a \emph{complete} instead of partial, wetting of NaCl(100) at the
melting point.

The increased temperature slope $|d\gamma^{*}_\mathrm{LV}/dT|$ confirms
that the calculated surface tension drop is directly related to the
restoring of a larger surface entropy, with removal of some of the SED through
cancellation of molecular surface correlations. We found in fact that the drop 
from $\gammaLV$ to
$\gamma^{*}_\mathrm{LV}$ at \Tmelt\ corresponds exactly to the increased
temperature slope $-d \gamma^{*}_\mathrm{LV}/dT$, that is to the surface entropy
increase, therefore with no change of surface internal energy as expected.
In the presence of the surface molecular short-range
order which we have just described, one could expect that the response to
an external electric field should be strongly reduced due to the
effective neutralization. Indeed this effect is present and very visible
in Heyes' early simulations of a BMHFT liquid slab in a parallel electric
field, which further confirms our interpretation~\cite{heyes}.

\section{Solid--liquid Interface Free Energy: Partial Wetting of a NaCl Droplet
On NaCl(100)}\label{sec:drop}

In the above Sections we calculated the SV and the LV surface free energies of
NaCl. The intermediate solid-liquid (SL) interface free energy
$\gamma_{SL}$, the third ingredient required to assess triple point wetting as 
in Eq.~(\ref{eq1}) is still missing.
We calculated $\gamma_{SL}$ through a simulation of the partial wetting
of solid NaCl(100) by a droplet of melt, and by using Young's equation 
Eq.~(\ref{mech}) that
connects it to the partial wetting angle~\cite{note3}.

\begin{figure}[!ht]
     \includegraphics[width=0.5\textwidth]{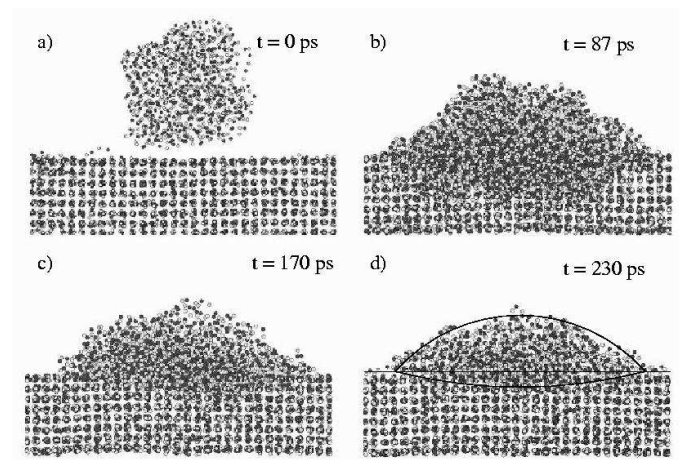}
     \caption{Time evolution of the liquid NaCl drop on NaCl(100).
     Dark and light circles stand for the Na$^+$ and Cl$^-$
     ions respectively.}
     \label{drop}
\end{figure}

A 500 molecule NaCl cluster was initially melted to form a nanodroplet. The
droplet and the solid slab were separately equilibrated at 1050~K, and then
brought to contact(fig.~\ref{drop}a). During the first 100 ps after contact, the 
droplet
settled onto the substrate, slightly spreading and gradually approaching a
final shape (fig.~\ref{drop}b). In the next 130 ps, spreading came to a
halt, and the settled liquid droplet survived in a metastable, long lived
state (Fig.~\ref{drop}c). At the end of the simulation, the droplet-substrate
system were as depicted in Fig.~\ref{drop}d.

Let us consider the thermodynamics of this situation. Because we are below
\Tmelt\ (even if slightly) the true final equilibrium state should
consist of a flat solid NaCl(100) surface, i.e.\ the nanodroplet should have
completely spread and recrystallized. That however will take a very long
time. While the nanodroplet still exists, it forms an external wetting angle
$\theta_{\mathrm{LV}}$ (Fig.~\ref{drop}), as well as an internal angle
$\theta_{\mathrm{SL}}$. These angles obey the mechanical equilibrium
equations Eq.~(\ref{mech})~\cite{nozieres}.
As it turns out, the angle $\theta_{\mathrm{SL}}$ is irrelevant here,
because it depends very critically on temperature. In particular close
enough to \Tmelt\ we expect $\theta_{\mathrm{SL}}\simeq 0$~\cite{nozieres}.
The external angle $\theta_{\mathrm{LV}}$ is instead fully significant,
as it depends very little on temperature, and indeed approximates
the macroscopic wetting angle measured in the bubble
experiment~\cite{mutaftschiev75,mutaftschiev97}.

To determine the external wetting angle $\theta_{\mathrm{LV}}$ of the
nanodroplet we analyzed 100 configurations in a 100 ps equilibrated run.
The instantaneous atomic positions were plotted in cylindrical coordinates
($r$ and $z$, where $r$ is parallel to the surface), and from the profile
of the drop, we determined the best approximation to a spherical segment,
by determining the center position and the radius.
The contact angle follows immediately by simple geometry from these two
quantities. Our best estimate obtained slightly below \Tmelt\ is
$\theta_\mathrm{LV} = 50^\circ \pm 5^\circ$. This value is in good
agreement with the experimental value at the melting point
(48$^\circ$)~\cite{mutaftschiev75}. At the end of the simulation the
internal solid-liquid interface was still relatively sharp and flat,
consistent with our assumption $\theta_{\mathrm{SL}}\simeq 0$.

The connection between $\theta_{\mathrm{LV}}$
and $\Delta\gamma_{\infty} = ( \gammaLV+\gammaSL ) - \gammaSV$
is given directly by Young's equation:
\begin{equation}
    \cos\theta_{\mathrm{LV}} = 1 -
      \frac{\Delta\gamma_{\infty}}{\gamma_{\mathrm{LV}}}.
\end{equation}
When we plug in our calculated value of $\gammaSV\simeq 103$ mJ/m$^2$, 
$\gammaLV\simeq 104$ mJ/m$^2$ and finally $\theta = 50^\circ \pm$ 5$^\circ$, we obtain
$\gammaSL = 36 \pm$ 6 mJ/m$^2$~\cite{notefrenkel} and $\Delta\gamma_{\infty} = 37$ mJ/m$^2$.

A solid-liquid interface free
energy of more than one-third the liquid surface tension is unusually large.
Here, it is clearly attributable to an unusually  large difference of density,
as well as of correlations, between the solid and the liquid at the melting
point. Corresponding to that, the solid-liquid interface is spatially
rather abrupt, as shown by Fig.~\ref{drop} and Fig.~\ref{sandwich}.

A further connection between $\Delta\gamma_{\infty}$ and the surface
spinodal temperature \Tss\ was made by assuming a phenomenological SL-LV
interface interaction of the form~\cite{ditolla95}:
\begin{equation}\label{eq:eq2}
     \Delta\gamma_{\infty}\simeq\rho L a \left(\frac{\Tss}{\Tmelt} - 1 \right).
\end{equation}
With our value of $\Delta\gamma_{\infty}$, and
$L$ = 4.813 $\cdot$ 10$^9$ erg/g, $a$ = 5.9~\AA\
we predict \Tss\ = 1210~K, quite close to that seen in simulations.

\section{Discussion and Conclusions}

We studied in this paper the physics of the solid NaCl(100) surface and
its wetting relationship with liquid NaCl at and near the melting
temperature of bulk NaCl. Molecular dynamics simulations performed with
classical BMHFT potentials were first of all shown to yield quite an
accurate description of high temperature solid and liquid bulk. The NaCl(100)
surface was subsequently studied, and found to be a non-melting surface,
one that should in principle be possible to overheat well above \Tmelt.
The solid surface free energy was calculated by thermodynamic integration
and seen to drop very considerably at high temperature due to large
anharmonicities. The liquid NaCl surface was also studied, and found to
be very diffuse, strongly fluctuating, and devoid of static structure such
as layering, or surface dipoles. However, calculation of the liquid surface
tension still gave a relatively large value, similar to the solid surface
free energy at the melting point. The high surface tension signifies an
unusual liquid surface entropy deficit, here ascribed
to short range molecular order. In addition, the solid liquid interface free
energy was also found to be relatively large, about one third the liquid surface
tension, consistent with the large density difference
between solid and liquid. Direct simulation of a NaCl droplet deposited
onto NaCl(100) demonstrated very realistically the incomplete wetting,
implied by the clear satisfaction of Eq.~(\ref{eq:young1}). At \Tmelt,
we obtained the solid, liquid and solid-liquid surface free energies, 
103$\pm$4, 104$\pm$8\,mJ/m$^2$, 36$\pm$6\,mJ/m$^2$ respectively, 
quantitatively explaining the surface nonmelting of NaCl(100) through
Eq.~(\ref{eq:young1}). We note that nonwetting of solid KCl by its own
liquid was arqued earlier by J. Rose and S. Berry~\cite{berry} based on the behavior
of a (KCl)$_{32}$ clusters.   

The present work is meant as a prototype study, ideally repeatable
for other elemental and molecular systems including perhaps water and the surface
of hexagonal ice~\cite{petrenko}.

In alkali halides, the extraordinarily poor wetting of the solid (100)
surface by the melt is traced  to the conspiracy of three separate factors, all 
of which can be
finally related to the long range Coulomb interaction between ions: (i)
surface anharmonicity stabilizes the solid surface; (ii) molecular
correlations destabilize the liquid surface; (iii) a large density jump
makes the solid-liquid interface very costly.

Several aspects uncovered in, or implied by, this study should be amenable
to direct experimental verification.
X-ray studies should confirm the diffuseness of the liquid
surface, the abruptness of the solid-liquid interface, and the molecular
short range order at the liquid surface.  Overheating of
solid NaCl(100) should be directly observable. So might the temporary
settling of liquid mini-droplets. Moreover, the much larger solid surface entropy
should cause the partial wetting angle of liquid on solid NaCl to
\emph{increase}, rather then decrease with temperature. A brief account of 
some of the present results has appeared in Ref.\cite{prl}.


\section{Acknowledgements}
Work in SISSA/ICTP/Democritos was sponsored by MIUR FIRB RBAU017S8R004,
FIRB RBAU01LX5H, and MIUR COFIN 2003, PRIN-COFIN2004, as well as by INFM
(section F,G, ``Iniziativa trasversale calcolo parallelo''). We acknowledge
illuminating discussions with E. A. Jagla and A. C. Levi, and the early
collaboration of W. Sekkal.



\begin{thebibliography}{99}
\bibitem{note1}
     A similar configuration is now predicted in NaCl by molecular dynamics
     simulation in Sec.~\ref{sec:drop} (see also Ref.~\cite{droplet}).
\bibitem{extrapolated}
     Strictly speaking the zero pressure melting temperature
     differs from the triple point temperature, which is slightly higher due
     to the finite triple point pressure. The triple point pressure is however
     only about 0.45 10$^{-6}$ kbar, which for $dP/dT \sim 0.03$ kbar/K yields
     extrapolated difference of only about 10$^{-5}$ K between the the triple
     point temperature and the zero-pressure melting temperature, utterly
     negligible.
\bibitem{note3}
     A preliminary account of partial wetting of NaCl(100)
     by a droplet of melt was presented in Ref~\cite{droplet}.
\end{thebibliography}
\end{document}